\documentclass[twocolumn,aps,prl,final,superscriptaddress,showpacs]{revtex4-1}
\usepackage[latin9]{inputenc}
\usepackage{color}
\usepackage{amsmath}
\usepackage{graphicx}
\usepackage[unicode=true,pdfusetitle,
 bookmarks=true,bookmarksnumbered=true,bookmarksopen=true,bookmarksopenlevel=1,
 breaklinks=true,pdfborder={0 0 1},backref=false,colorlinks=true]
 {hyperref}
\hypersetup{
 linkcolor=blue,urlcolor=blue,citecolor=blue,pdfstartview={FitH},hyperfootnotes=false}

\makeatletter
\usepackage{dcolumn}
\usepackage{bm}

\usepackage{times}

\makeatother

\begin{document}
\title{Electrocaloric effects in multiferroics}
\author{Zhijun Jiang}
\affiliation{MOE Key Laboratory for Nonequilibrium Synthesis and Modulation of
Condensed Matter, School of Physics, Xi'an Jiaotong University, Xi'an
710049, China}
\affiliation{Physics Department and Institute for Nanoscience and Engineering,
University of Arkansas, Fayetteville, Arkansas 72701, USA}
\affiliation{Key Laboratory of Computational Physical Sciences (Ministry of Education),
State Key Laboratory of Surface Physics, and Department of Physics,
Fudan University, Shanghai 200433, China}
\author{Bin Xu}
\affiliation{Jiangsu Key Laboratory of Thin Films, School of Physical Science and
Technology, Soochow University, Suzhou 215006, China}
\author{Sergey Prosandeev}
\affiliation{Physics Department and Institute for Nanoscience and Engineering,
University of Arkansas, Fayetteville, Arkansas 72701, USA}
\author{Yousra Nahas}
\affiliation{Physics Department and Institute for Nanoscience and Engineering,
University of Arkansas, Fayetteville, Arkansas 72701, USA}
\author{Sergei Prokhorenko}
\affiliation{Physics Department and Institute for Nanoscience and Engineering,
University of Arkansas, Fayetteville, Arkansas 72701, USA}
\author{Jorge Íñiguez}
\affiliation{Materials Research and Technology Department, Luxembourg Institute
of Science and Technology, 5 Avenue des Hauts-Fourneaux, L-4362, Esch/Alzette,
Luxembourg}
\affiliation{Physics and Materials Science Research Unit, University of Luxembourg,
41 Rue du Brill, L-4422 Belvaux, Luxembourg}
\author{L. Bellaiche}
\affiliation{Physics Department and Institute for Nanoscience and Engineering,
University of Arkansas, Fayetteville, Arkansas 72701, USA}
\begin{abstract}
An atomistic effective Hamiltonian is used to compute electrocaloric
(EC) effects in rare-earth substituted BiFeO$_{3}$ multiferroics.
A phenomenological model is then developed to interpret these computations,
with this model indicating that the EC coefficient is the sum of two
terms, that involve electric quantities (polarization, dielectric
response), the antiferromagnetic order parameter, and the coupling
between polarization and antiferromagnetic order. The first one depends
on the polarization and dielectric susceptibility, has the analytical
form previously demonstrated for ferroelectrics, and is thus enhanced
at the ferroelectric Curie temperature. The second one explicitly
involves the dielectric response, the magnetic order parameter and
a specific magnetoelectric coupling, and generates a peak of the EC
response at the Néel temperature. These atomistic results and phenomenological
model may be put in use to optimize EC coefficients. 
\end{abstract}
\maketitle
The electrocaloric (EC) effect is a phenomenon by which a material
exhibits a reversible temperature change under the application/removal
of an electric field \cite{Lines1997,Scott2007,Scott2011,Zhang2014,Kutnjak2015}.
It is attracting attention due to its potential to be an efficient
solid-state refrigeration technology (see, e.g., Refs.\ \cite{Uchino2000,Zhang2006,Prosandeev2008,Ponomareva2012,Rose2012,Defay2013,Moya2014,Geng2015,Marathe2016,Guzman-Verri2016,Jiang2017,Jiang2018,Nair2019,Shi2019}
and references therein).

Furthermore, multicaloric effects that are driven simultaneously by
more than one type of external physical handle, such as electric and/or
magnetic fields, mechanical stress and pressure \cite{Vopson2012,Stern-Taulats2018,Liu2016,Takeuchi2015,Khassaf2017},
are also promising to enhance change in temperature \cite{Takeuchi2015,Khassaf2017}.

Recently, multiferroics, which are materials that possess coupled
long-range-ordered electric and magnetic degrees of freedom \cite{Catalan2009,Zhao2006,Lebeugle2008,Zeches2009,Spaldin2010,Xu2017,Spaldin2019},
have also been mentioned as possible systems to enhance the EC effects
by taking advantage of such coupling \cite{Moya2014,Vopson2012,Stern-Taulats2018,Cazorla2018,Edstrom2019,Zhao2020}.
The pioneering work of Ref.\ \cite{Edstrom2019} started from a phenomenological
Landau-type equation for which coefficients were determined from first
principles to investigate how magnetoelectric coupling modifies the
EC coefficient. The main result was that EC effects are significantly
enhanced (by about $60\%$) thanks to magnetoelectric coupling in
the case that the ferroelectric and magnetic critical temperatures
coincide. However, one has to be careful when using a Landau-type
approach because fluctuations, which can be important for responses,
are not treated explicitly and may be underestimated. That is why
atomistic approaches incorporating couplings between electric dipoles
and spins can be useful to also study EC effects in multiferroics,
as the authors of Ref.\ \cite{Edstrom2019} indicated. More importantly,
it is presently unclear how to understand EC coefficients in multiferroics.
For instance, can these coefficients be considered as composed of
two terms, with one corresponding to that occurring in normal ferroelectrics
and the second one related to the coupling between spins and electric
dipoles? If yes, what are the precise quantities involved in the second
term? Are they only magnetoelectric, or rather also involve electric
and/or magnetic properties? Answering such questions will help in
designing systems with large EC response.

The aim of this Letter is to resolve all these issues by (1) conducting
atomistic-based simulations; (2) developing a simple model that can
reproduce these simulations; and (3) using such simulations and model
to gain a deep microscopic insight. We demonstrate that the EC coefficient
of multiferroics can be thought as having two parts, each associated
with different physical quantities.

Here, we adopt the effective Hamiltonian ($H_{\mathrm{eff}}$) approach
developed in Ref.\ \cite{Xu2015} to study disordered Bi$_{1-x}$Nd$_{x}$FeO$_{3}$
(BNFO) alloys. $H_{\mathrm{eff}}$ parameters are provided in the
Supplemental Material (SM) \cite{Supplemental Material}. This $H_{\mathrm{eff}}$
successfully reproduced the temperature-\textit{versus}-compositional
phase diagram of BNFO. It predicts a $R3c$ ground state for small
Nd compositions and a $Pnma$ phase for larger concentrations, with
intermediate complex states in-between. Moreover, within the compositional
range for which the $R3c$ phase is the ground state, the ferroelectric
Curie temperature $T_{\mathrm{C}}$ was numerically found to significantly
decrease with the Nd composition while the $T_{\mathrm{N}}$ Néel
temperature is mostly independent of concentration, which also agrees
with measurements \cite{Karimi2009,Levin2010,Levin2011}. The total
internal energy of this $H_{\mathrm{eff}}$ can be expressed as a
sum of two main terms:

\begin{eqnarray}
E_{\mathrm{tot}} & = & E_{\mathrm{BFO}}(\{\mathrm{\mathbf{u}}_{i}\},\thinspace\{\eta_{\mathrm{H}}\},\thinspace\{\eta_{\mathrm{I}}\},\thinspace\{\mathbf{\boldsymbol{\omega}}_{i}\},\thinspace\{\mathrm{\mathbf{m}}_{i}\})\nonumber \\
 &  & +\thinspace E_{\mathrm{alloy}}(\{\mathrm{\mathbf{u}}_{i}\},\thinspace\{\mathbf{\boldsymbol{\omega}}_{i}\},\thinspace\{\mathrm{\mathbf{m}}_{i}\},\thinspace\{\eta_{\mathrm{loc}}\})\thinspace,\label{eq:total_energy}
\end{eqnarray}
where $E_{\mathrm{BFO}}$ is the $H_{\mathrm{eff}}$ of pure BiFeO$_{3}$
\cite{Kornev2007,Lisenkov2009,Albrecht2010,Prosandeev2013-AFM} and
$E_{\mathrm{alloy}}$ characterizes the effect of substituting Bi
by Nd ions. The $H_{\mathrm{eff}}$ of BNFO contains four types of
degrees of freedom: (i) the local soft mode $\{\mathrm{\mathbf{u}}_{i}\}$
centered on the A site of Bi or Nd ions in the $5$-atom unit cell
$i$ (which is proportional to the local electric dipole moment of
that cell \cite{Zhong1994,Zhong1995}); (ii) the strain tensor gathering
homogeneous $\{\eta_{\mathrm{H}}\}$ and inhomogeneous $\{\eta_{\mathrm{I}}\}$
contributions \cite{Zhong1994,Zhong1995}; (iii) the pseudovectors
$\{\mathbf{\boldsymbol{\omega}}_{i}\}$ that represent the oxygen
octahedral tiltings \cite{Kornev2006}; and (iv) the magnetic moments
$\{\mathrm{\mathbf{m}}_{i}\}$ centered on Fe ions \cite{note-1}.

We employ this $H_{\mathrm{eff}}$ within Monte Carlo (MC) simulations
on $12\times12\times12$ supercells (containing $8~640$ atoms) with
periodic boundary conditions and inside which Bi and Nd ions are randomly
distributed over the A sublattice. $20~000$ MC sweeps are used for
equilibration and an additional $20~000$ MC sweeps are employed to
compute statistical averages at finite temperature, to obtain converged
results. We also average our results over $10$ random Bi/Nd distributions,
to mimic well disordered BNFO solid solutions.

Regarding the linear EC coefficient, $\alpha_{\gamma}$, it is the
derivative of the temperature with respect to electric field at constant
entropy. It can be obtained from MC simulations by taking advantage
of the cumulant formula \cite{Jiang2017,Jiang2018,Omran2016}: 
\begin{equation}
\alpha_{\gamma}=-\thinspace Z^{*}a_{\mathrm{lat}}T\thinspace\left\{ \frac{\left\langle u_{\gamma}{E_{\mathrm{tot}}}\right\rangle -\left\langle u_{\gamma}\right\rangle \left\langle {E_{\mathrm{tot}}}\right\rangle }{\left\langle {E_{\mathrm{tot}}}^{2}\right\rangle -\left\langle {E_{\mathrm{tot}}}\right\rangle ^{2}+\frac{21(k_{B}T)^{2}}{2N}}\right\} \thinspace,\label{eq:ece-alpha-cumulant}
\end{equation}
where $Z^{*}$ is the Born effective charge associated with the local
mode, $a_{\mathrm{lat}}$ represents the five-atom lattice constant,
$T$ is the temperature, $u_{\gamma}$ is the $\gamma$-component
of the supercell average of the local mode with $\gamma=x$, $y$,
or $z$ (note that the $x$, $y$, and $z$ axis are chosen along
the pseudocubic $[100]$, $[010]$ and $[001]$ directions, respectively),
${E_{\mathrm{tot}}}$ is the total internal energy given by the $H_{\mathrm{eff}}$,
$k_{B}$ is the Boltzmann constant, $N$ is the number of sites in
the supercell, and $\left\langle \ \right\rangle $ defines the average
over the $\textrm{MC}$ sweeps at a given temperature \cite{note-2}.
In the following, we will denote $\alpha$ the quantity defined by
$\frac{\alpha_{x}+\alpha_{y}+\alpha_{z}}{\sqrt{3}}$. Such definition
corresponds to the EC response for an electric field applied along
$[111]$, which is the maximal response within a $R3c$ state.

\begin{figure}
\includegraphics[width=7.5cm]{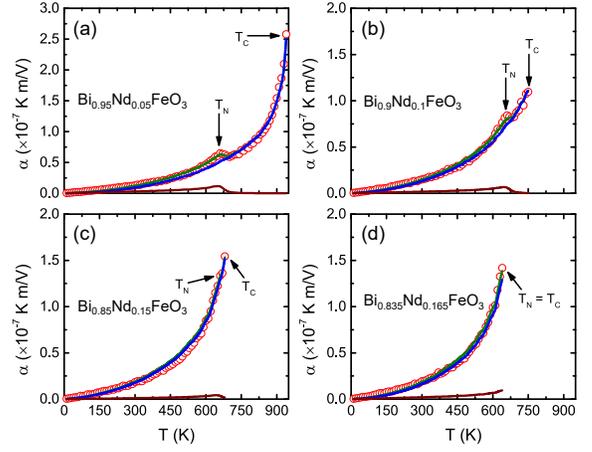}

\caption{Electrocaloric coefficient, $\alpha$, as a function of the temperature
for different compositions in disordered Bi$_{1-x}$Nd$_{x}$FeO$_{3}$
alloys: (a) Bi$_{0.95}$Nd$_{0.05}$FeO$_{3}$; (b) Bi$_{0.9}$Nd$_{0.1}$FeO$_{3}$;
(c) Bi$_{0.85}$Nd$_{0.15}$FeO$_{3}$; and (d) Bi$_{0.835}$Nd$_{0.165}$FeO$_{3}$.
The solid green lines represent the fit of the $\textrm{MC}$ results
by the second line of Eq.\ (\ref{eq:Landau-like-model}), i.e., $\alpha=\frac{T_{0}a'(T)}{C_{ph}}P_{s}\varepsilon_{0}\chi+\frac{T_{0}b'(T)}{C_{ph}}L_{s}\left.\frac{\partial L_{s}}{\partial P_{s}}\right|_{T}\varepsilon_{0}\chi$,
where $a'(T)=A_{0}+A_{1}T$ ($A_{0}$ and $A_{1}$ being fitting constants),
and $C_{ph}$ and $b'(T)$ are also fitting parameters. The solid
blue lines display the fit of the MC results by its first contribution,
$\frac{T_{0}a'(T)}{C_{ph}}P_{s}\varepsilon_{0}\chi$. The solid brown
lines correspond to the fit of the MC results by its second contribution,
$\frac{T_{0}b'(T)}{C_{ph}}L_{s}\left.\frac{\partial L_{s}}{\partial P_{s}}\right|_{T}\varepsilon_{0}\chi$
(see text). \label{fig:alpha_vs_T}}
\end{figure}

Figure\ \ref{fig:alpha_vs_T} shows the EC coefficient as a function
of temperature for four different Nd compositions in disordered Bi$_{1-x}$Nd$_{x}$FeO$_{3}$.
The results of Fig.\ \ref{fig:alpha_vs_T} are obtained by starting
from $10~\textrm{K}$ adopting a $R3c$ phase and then progressively
heating up the BNFO solid solutions up to the composition-dependent
Curie temperature, $T_{\mathrm{C}}$ (for all investigated temperatures
displayed in Fig.\ \ref{fig:alpha_vs_T}, the disordered Bi$_{1-x}$Nd$_{x}$FeO$_{3}$
alloys possess the $R3c$ phase from $0~\textrm{K}$ and up to $T_{\mathrm{C}}$).
This $R3c$ state is characterized by a polarization lying along $[111]$
and oxygen octahedra tilting in an antiphase fashion about this polarization's
direction. These solid solutions also exhibit a G-type antiferromagnetic-to-paramagnetic
transition at a Néel temperature, $T_{\mathrm{N}}$, which is mostly
independent on the composition and equal to $\simeq660~\textrm{K}$
\cite{Xu2015}. The SM \cite{Supplemental Material} provides some
finite-temperature properties above $T_{\mathrm{C}}$.

Let us first focus on Fig.\ \ref{fig:alpha_vs_T}(a) that corresponds
to a concentration of Nd equal to $5\%$. The calculated $T_{\mathrm{C}}\simeq940~\textrm{K}$
and $T_{\mathrm{N}}\simeq660~\textrm{K}$ of Bi$_{0.95}$Nd$_{0.05}$FeO$_{3}$
are in rather good agreement with the measurements of $T_{\mathrm{C}}\simeq970~\textrm{K}$
and $T_{\mathrm{N}}\simeq650~\textrm{K}$ \cite{Levin2010,Levin2011}.
For any investigated temperature, $\alpha$ basically monotonically
increases when the system is heated up to the Néel temperature. It
then adopts a small peak around $T_{\mathrm{N}}$, which is found
to originate from the coupling between polarization and magnetism
-- we verify this by running $H_{\mathrm{eff}}$ simulations in which
the coupling between local models and magnetic moments is turned off.
The EC coefficient then significantly strengthens when increasing
the temperature from the end of this $\simeq$ $T_{\mathrm{N}}$-centered
peak and up to $T_{\mathrm{C}}$. Our predicted big value of $\alpha$
around $T_{\mathrm{C}}$ is of the order of $\simeq2.6\times10^{-7}$
$\textrm{K m/V}$. It is thus large and close to the experimental
data of $2.5\times10^{-7}$ $\textrm{K m/V}$ at $T\simeq499~\textrm{K}$
in PbZr$_{0.95}$Ti$_{0.05}$O$_{3}$ films \cite{Zhang2006} (the
largest observed $\alpha$ is equal to $22\times10^{-7}$ $\textrm{K m/V}$
and has been found in a BaTiO$_{3}$ single crystal, see Ref.\ \cite{Moya2013})
\cite{footnote-1}. Note that $H_{\mathrm{eff}}$ techniques have
been demonstrated in Refs.\ \cite{Jiang2017,Jiang2018} to accurately
reproduce the EC coefficients of ferroelectrics and relaxor ferroelectrics,
such as those reported in BaTiO$_{3}$ \cite{Karchevskii1962,Moya2013}
and Pb(Mg,Nb)O$_{3}$ \cite{Rozic2011}.

Let us now concentrate on other compositions in disordered Bi$_{1-x}$Nd$_{x}$FeO$_{3}$
alloys. Figures\ \ref{fig:alpha_vs_T}(b)-\ref{fig:alpha_vs_T}(d)
show the dependence of the EC coefficient when the Nd composition
is equal to $x$ $=$$0.10$, $0.15$ and $0.165$, respectively.
The Curie temperature $T_{\mathrm{C}}$ noticeably decreases when
increasing the Nd composition, as consistent with observations and
computations \cite{Xu2015,Karimi2009,Levin2010,Levin2011}. Consequently,
the two critical temperatures coincide, i.e. $T_{\mathrm{C}}=T_{\mathrm{N}}$,
for a Nd concentration of $16.5\%$. Figures\ \ref{fig:alpha_vs_T}(b)-\ref{fig:alpha_vs_T}(d)
especially reveals that $\alpha$ at the Néel temperature is enhanced
when the Nd composition increases, but it becomes more difficult to
see its associated peak.

To understand the results in Fig.\ \ref{fig:alpha_vs_T}, we use
a Landau free-energy potential $F(P,L,{\cal E},T)$ in which we substitute
polarization $P$ and G-type antiferromagnetic (AFM) moment $L$ by
their equilibrium values $P_{s}$ and $L_{s}$ found from minimization
of free energy: $\left.\frac{\partial F}{\partial P}\right|_{P=P_{s},{\cal E},T}=0$
and $\left.\frac{\partial F}{\partial L}\right|_{L=L_{s},{\cal E},T}=0$.
The minimized free energy $F_{s}({\cal E},T)=F(P_{s},L_{s},{\cal E},T)$
has the form:

\begin{equation}
\begin{split}F_{s}({\cal E},T) & =\frac{1}{2}a(T)P_{s}^{\mathrm{2}}({\cal E},T)+\frac{1}{4}\beta P_{s}^{\mathrm{4}}({\cal E},T)\\
 & \quad-{\cal E}P_{s}({\cal E},T)+\frac{1}{2}b(T)L_{s}^{\mathrm{2}}({\cal E},T)\\
 & \quad+\frac{1}{4}\kappa L_{s}^{\mathrm{4}}({\cal E},T)+\frac{1}{2}cL_{s}^{\mathrm{2}}({\cal E},T)P_{s}^{\mathrm{2}}({\cal E},T)\thinspace,
\end{split}
\label{eq:Landau expansion}
\end{equation}
where ${\cal E}$ is the electric field.

Such equation implies that the polarization implicitly depends on
magnetism, because of the $\frac{1}{2}cL_{s}^{\mathrm{2}}({\cal E},T)P_{s}^{\mathrm{2}}({\cal E},T)$
term. This equation is similar to the one used in Ref.\ \cite{Edstrom2019}.
The entropy described by this free energy $F_{s}({\cal E},T)$, composed
of dipoles and spins, can then be obtained as

\begin{equation}
\begin{split}S_{F}({\cal E},T) & =-\left.\frac{dF_{s}}{dT}\right|_{{\cal E}}\\
 & =-\frac{a'(T)}{2}P_{s}^{\mathrm{2}}({\cal E},T)-\frac{b'(T)}{2}L_{s}^{\mathrm{2}}({\cal E},T)\thinspace\text{,}
\end{split}
\label{eq:entropy}
\end{equation}
where $a'=da/dT$ and $b'=db/dT$. Note that, here we took into account
that $P_{s}$ and $L_{s}$ are found from minimization of the free
energy.

In the case of a magnetic phase transition and presence of polarization,
we can consider two parts of the \textit{total} entropy $S({\cal E},T)$:
A first one due to electric dipoles and spins (the \textsl{active}
part treated by the Landau potential above, with entropy $S_{F}({\cal E},T)$)
and a second one due to the rest of the lattice (the inert part that
can be considered to be a trivial collection of harmonic phonons,
with entropy $S_{ph}(T)$) \cite{Pirc2011,Pirc2014}. For an adiabatic
process, we have: 
\begin{equation}
\Delta S({\cal E},T)=\Delta S_{F}({\cal E},T)+\Delta S_{ph}(T)=0\thinspace.\label{eq:the change of entropy}
\end{equation}
Let $C_{ph}$ denote the heat capacity associated with the background
lattice modes. Then the change of lattice entropy from an initial
state $(0,T_{0})$ to the final state $({\cal E},T)$ is given by:

\begin{equation}
\Delta S_{ph}=\int_{T_{0}}^{T}\frac{C_{ph}}{T}dT\cong C_{ph}\ln\left(\frac{T}{T_{0}}\right)\thinspace.\label{eq:the change of lattice entropy}
\end{equation}
Consequently, combining Eqs.\ (\ref{eq:the change of entropy}) and
(\ref{eq:the change of lattice entropy}) leads to 
\begin{equation}
C_{ph}\ln\left(\frac{T}{T_{0}}\right)=-\Delta S_{F}=\frac{1}{2}a'(P_{s}^{\mathrm{2}}-P_{0}^{2})+\frac{1}{2}b'(L_{s}^{\mathrm{2}}-L_{0}^{2})\thinspace.
\end{equation}
Here $P_{s}=P_{s}({\cal E},T)$, $P_{0}=P_{s}(0,T_{0})$, $L_{s}=L_{s}({\cal E},T)$,
$L_{0}=L_{s}(0,T_{0})$, where $T_{0}$ is the initial temperature
and $T=T_{0}+\Delta T$ is the final temperature ($\Delta T$ represents
the temperature change). Solving this equation with respect to $T/T_{0}$
yields: 
\begin{equation}
(T_{0}+\Delta T)/T_{0}=e^{\left[a'(P_{s}^{\mathrm{2}}-P_{0}^{2})+b'(L_{s}^{\mathrm{2}}-L_{0}^{2})\right]/2C_{ph}}\thinspace.
\end{equation}
For small $\Delta T$: 
\begin{equation}
\Delta T=\frac{T_{0}\left[a'(P_{s}^{\mathrm{2}}-P_{0}^{2})+b'(L_{s}^{\mathrm{2}}-L_{0}^{2})\right]}{2C_{ph}}\thinspace.\label{eq:deltaT}
\end{equation}

One can then derive the following expression for $\alpha$ \cite{Rose2012,Jiang2017}:

\begin{equation}
\alpha=\left.\frac{\partial\Delta T}{\partial{\cal E}}\right|_{S}\approx\left.\frac{T_{0}a'(T)}{2C_{ph}}\frac{\partial P_{s}^{\mathrm{2}}}{\partial{\cal E}}\right|_{T}+\left.\frac{T_{0}b'(T)}{2C_{ph}}\frac{\partial L_{s}^{\mathrm{2}}}{\partial{\cal E}}\right|_{T}\thinspace.
\end{equation}

Here we assumed that, since the adiabatic temperature change is small
as compared to the temperature, the constant-$S$ derivatives can
be evaluated at a constant $T=T_{0}$. One can write:

\begin{equation}
\begin{split}\alpha & =\frac{T_{0}a'(T)}{C_{ph}}P_{s}\varepsilon_{0}\chi+\frac{T_{0}b'(T)}{2C_{ph}}\left.\frac{\partial L_{s}^{\mathrm{2}}}{\partial P_{s}}\right|_{T}\left.\frac{\partial P_{s}}{\partial{\cal E}}\right|_{T}\\
 & =\frac{T_{0}a'(T)}{C_{ph}}P_{s}\varepsilon_{0}\chi+\frac{T_{0}b'(T)}{C_{ph}}L_{s}\left.\frac{\partial L_{s}}{\partial P_{s}}\right|_{T}\varepsilon_{0}\chi\thinspace,
\end{split}
\label{eq:Landau-like-model}
\end{equation}
where $\varepsilon_{0}$ is the vacuum permittivity and $\chi$ is
the dielectric susceptibility. Finally, let us note that one could
try to approximate $C_{ph}$ by adding a $k_{B}$ contribution for
each degree of freedom belonging to the trivial -- harmonic -- part
of the system. However, it is not obvious how to count the exact number
of active and inactive variables in the framework of a Landau theory;
we thus treat $C_{ph}$ as an adjustable parameter. Note that we did
not fit $C_{ph}$ alone but rather the ratio of $a'(T)$/$C_{ph}$
and $b'(T)$/$C_{ph}$.

\begin{figure}
\includegraphics[width=7.5cm]{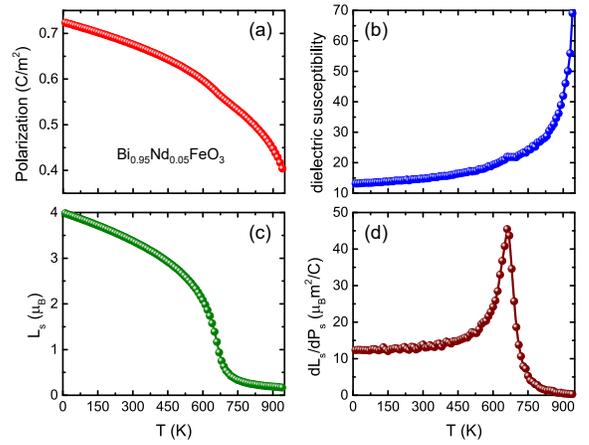}

\caption{Temperature dependence of some properties in disordered Bi$_{0.95}$Nd$_{0.05}$FeO$_{3}$
alloys, as obtained from our MC simulations: (a) the macroscopic polarization
$P_{s}$; (b) the average between the three diagonal elements of the
dielectric susceptibility; (c) the AFM vector; and (d) the derivative
$dL_{s}$/$dP_{s}$. \label{fig:properties versus T}}
\end{figure}

As shown by the green lines of Fig.\ \ref{fig:alpha_vs_T}, the second
line of Eq.\ (\ref{eq:Landau-like-model}) fits well the MC data,
when (1) using the $P_{s}$, $\chi$, $L_{s}$ and $\frac{\partial L_{s}}{\partial P_{s}}$
\cite{dL/dP} obtained by our Monte-Carlo simulations (these four
quantities are shown in Fig.\ \ref{fig:properties versus T} for
the case of a $5\%$ Nd composition); and (2) assuming that $C_{ph}$
and $b'(T)$ are fitting constants, while $a'(T)=A_{0}+A_{1}T$ with
$A_{0}$ and $A_{1}$ are fitting parameters \cite{note-3}. Since
its validity is confirmed by Fig.\ \ref{fig:alpha_vs_T}, the second
line of Eq.\ (\ref{eq:Landau-like-model}) can now be used to gain
an insight \cite{Supplemental Material} into the results of Fig.\ \ref{fig:alpha_vs_T},
via the decomposition of $\alpha$ into its two terms -- that are
$\frac{T_{0}a'(T)}{C_{ph}}P_{s}\varepsilon_{0}\chi$ and $\frac{T_{0}b'(T)}{C_{ph}}L_{s}\left.\frac{\partial L_{s}}{\partial P_{s}}\right|_{T}\varepsilon_{0}\chi$.
The first contribution has precisely the analytical form of the EC
coefficient for \textit{non-magnetic systems}, see Refs.\ \cite{Jiang2017,Jiang2018}.
It is shown by blue lines in Fig.\ \ref{fig:alpha_vs_T}, and is
the one that contributes the most to the total $\alpha$ for any composition.
Its increases with temperature and is driven by the corresponding
increase in dielectric susceptibility, however moderated by the concomitant
decrease in polarization {[}see Figs.\ \ref{fig:properties versus T}(b)
and \ref{fig:properties versus T}(a){]}. This first contribution
implicitly depends on magnetism because of the coupling between polarization
and antiferromagnetism, as evidenced in the change of behavior of
the polarization and in the occurrence of a plateau in the dielectric
response near $T_{\mathrm{N}}$ (such behavior of $\chi$ has been
reported in other multiferroics \cite{Kornev2007,Tomuta2001}). The
second contribution of Eq.\ (\ref{eq:Landau-like-model}) is depicted
in brown lines in Fig.\ \ref{fig:alpha_vs_T}, and is basically independent
on the investigated composition for any temperature. As evidenced
in Fig.\ \ref{fig:alpha_vs_T}, it is the one responsible for the
small peak of $\alpha$ found near the Néel temperature. This small
peak becomes more difficult to be seen in the total EC coefficient
(shown in green) when the Nd composition increases simply because
the first contribution provides much larger values than the second
contribution. Figures\ \ref{fig:properties versus T}(c) and \ref{fig:properties versus T}(d)
also reveal that this small peak originates from the activation and
then sharp increase of the magnitude of $\frac{\partial L_{s}}{\partial P_{s}}$
near $T_{\mathrm{N}}$. This derivative for temperatures far away
below $T_{\mathrm{N}}$ is then basically a constant that characterizes
intrinsic magnetoelectric coupling -- which is related to the $c$
constant of Eq.\ (\ref{eq:Landau expansion}). The second term of
Eq.\ (\ref{eq:Landau-like-model}) tells us that the EC coefficient
of a multiferroic can be optimized even at temperatures far away $T_{\mathrm{N}}$
in systems possessing strong coupling between polarization and magnetic
ordering. Ba(Sr,Ba)MnO$_{3}$ films may thus be a system of choice
to investigate electrocaloric effects due to its strong magnetoelectric
coupling \cite{Bayaraa2018,Sakai2011,Maurel2019}.

\begin{figure}
\includegraphics[width=7cm]{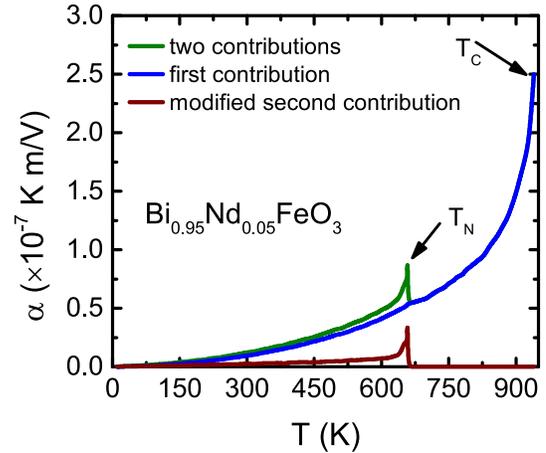}

\caption{Same as Fig. \ref{fig:alpha_vs_T} (a) but now using a different $\frac{\partial L_{s}}{\partial P_{s}}$
(see text) in the second line of Eq.\ (\ref{eq:Landau-like-model}).
\label{fig:alpha_change_AFM}}
\end{figure}

The now-elucidated effect of $\frac{\partial L_{s}}{\partial P_{s}}$
on $\alpha$ near $T_{\mathrm{N}}$ can be further used to address
the finite-size effects in our computations of the EC coefficient.
It is known that such size effect broadens the magnetic transition
when decreasing the supercell size (see the SM \cite{Supplemental Material})
\cite{Parnaste2005,Mokkath2020}, and we also checked that the magnitude
of the second contribution of $\alpha$ around $T_{\mathrm{N}}$ increases
when increasing such size. It will thus be more realistic, regarding
what to expect in experiments, to rather adopt a $L_{s}=A|T_{\mathrm{N}}-T|^{\beta}$
power law (see Refs.\ \cite{Parnaste2005,Ashcroft1976}) near the
Néel temperature, where $A$ and $\beta$ are coefficients. Consequently,
we (1) chose to replace, around $T_{\mathrm{N}}$, the MC data for
$L_{s}$ by the result given by such power law with $\beta$ equal
to $0.5$ (mean-field value); (2) continue to still use the MC data
for $L_{s}$ for temperatures far away (below) the Néel temperature;
and (3) extract $A$ such by imposing that this power law of item
(1) matches the MC data of item (2). Using the new resulting $\frac{\partial L_{s}}{\partial P_{s}}$
along with all the previous other quantities in Eq.\ (\ref{eq:Landau-like-model})
(including the temperature behavior of the polarization) provides
the data given in Fig.\ \ref{fig:alpha_change_AFM} for the second
contribution but also total EC coefficient in disordered Bi$_{0.95}$Nd$_{0.05}$FeO$_{3}$
alloys. The aforementioned change of $L_{s}$'s behavior, that is
a more abrupt change near $T_{\mathrm{N}}$, leads to a narrower and
stronger peak of $\alpha$ close to the Néel temperature. The second
contribution now amounts for $42\%$ of the total EC coefficient near
the magnetic transition. Such latter result is in-line with the phenomenological
theory of Edström \textit{et al.} \cite{Edstrom2019} predicting that
the magnetic contribution can reach approximately $60\%$ of the electric
contribution at the magnetic transition, and thus enhance the EC effect,
in epitaxial multiferroic SrMnO$_{3}$ systems under a tensile strain
of $2.63\%$ -- for which $T_{\mathrm{N}}=T_{\mathrm{C}}$. Our study
explains why it is the case thanks to Eq.\ (\ref{eq:Landau-like-model})
that not only reproduces atomistic results but also and especially
provides an insight into the microscopic origins of the EC effects
in a multiferroic. We also used a larger supercell and such power
law of $L_{s}$ with different $\beta$, and found that our qualitative
results are still valid for any reasonable choice of $\beta$ (see
Fig.\ S3 of the SM \cite{Supplemental Material}). Note that the
peak of Fig.\ \ref{fig:alpha_vs_T}(a) at the Néel temperature is
significantly less pronounced than in Ref.\ \cite{Edstrom2019} for
two possible reasons. The first one is that such peak depends on the
size of the simulation supercell (see the SM \cite{Supplemental Material})
and the second one is that the magnetoelectric coupling is weaker
in BiFeO$_{3}$ \cite{Kornev2007} than in SrMnO$_{3}$ \cite{Edstrom2019}.
Fluctuations within the $H_{\mathrm{eff}}$ are also discussed in
the SM \cite{Supplemental Material}.

In summary, an atomistic effective Hamiltonian scheme has been used
to compute finite-temperature electrocaloric coefficients in the rare-earth
substituted BiFeO$_{3}$ multiferroic. The results are then interpreted
via the development of a model that reproduces these computational
data. EC coefficients can be decomposed in two main terms. The first
term takes its largest value at the Curie temperature and explicitly
depends on the polarization and dielectric susceptibility, that are
both implicit functions of magnetic ordering and strength because
of magnetoelectric couplings. The second term adopts a peak near the
Néel temperature and is proportional to the antiferromagnetic vector,
the polarization derivative of the antiferromagnetic vector and the
dielectric susceptibility. Such findings therefore suggest an original
way to induce large EC coefficients by simultaneous optimization of
electric, magnetic and magnetoelectric properties at a selected temperature
below the Néel temperature: (1) the dielectric susceptibility should
be large; (2) the antiferromagnetic vector should be strong; and (3)
the magnetoelectric coupling $\frac{\partial L_{s}}{\partial P_{s}}$
should be large \cite{footnote-2}. Our results and phenomenology
should be valid for all magnetoelectric multiferroics, at the exception
of those for which a magnetic Dzyaloshinskii-Moriya interaction involving
the polarization (e.g., the spin-current model) is important. We hope
that the present article deepens the fields of multiferroics and important
subtle cross-coupling properties such as electrocaloric effects. 
\begin{acknowledgments}
This work is supported by the National Natural Science Foundation
of China (Grants No.\ 11804138 and No.\ 11825403), Shandong Provincial
Natural Science Foundation (Grant No.\ ZR2019QA008), China Postdoctoral
Science Foundation (Grants No.\ 2020T130120 and No.\ 2018M641905),
``Young Talent Support Plan'' of Xi'an Jiaotong University, Postdoctoral
International Exchange Program of Academic Exchange Project, and Shanghai
Post-doctoral Excellence Program. B.\ X.\ acknowledges financial
support from National Natural Science Foundation of China (Grant No.\ 12074277),
the startup fund from Soochow University and support from Priority
Academic Program Development (PAPD) of Jiangsu Higher Education Institutions.
S.\ Prosandeev is supported by ONR Grant N00014-17-1-2818. Y.\ N.,
S.\ Prokhorenko and L.\ B. thank the DARPA Grants No. HR0011727183-D18AP00010
(TEE programme) and No. HR0011-15-2-0038 (MATRIX program). J.\ Í.
acknowledges funding from the Luxembourg National Research Fund through
the CORE program (Grant No. FNR/C18/MS/12705883 REFOX, J.\ Í.). 
\end{acknowledgments}


\begin{thebibliography}{10}
\bibitem{note-1}Note that the local quantity $\eta_{\mathrm{loc}}(i)$
is centered on the Fe-site $i$ and is defined as $\eta_{\mathrm{loc}}(i)$=$\frac{\delta R_{\mathrm{ionic}}}{8}\sum_{j}\sigma_{j}$,
where $\sigma_{j}$ characterizes the atomic distribution of Bi or
Nd ion at the A site $j$ and the sum over $j$ runs over the eight
A nearest neighbors of the Fe-site $i$. $\delta R_{\mathrm{ionic}}$
describes the relative difference of ionic radius between the Nd and
Bi ions within the A sublattice.
\bibitem{note-2}Note that the denominator in this expression essentially
corresponds to the calculated specific heat, which has several parts.
The non-trivial one is associated from the potential energy of the
subsystem described by our effective Hamiltonian (local dipoles, O$_{6}$
rotations, inhomogeneous strains and spins), and is given by the width
of the energy distribution as obtained from our Monte Carlo simulations.
The second, trivial part is associated to the kinetic contribution
of all the lattice degrees of freedom in our material (15 per cell),
as well as the contribution associated to the potential energy of
the variables not included in our effective Hamiltonian (6 per cell);
hence the term proportional to $21k_{B}T/2$. Note also that Eq.\ (\ref{eq:ece-alpha-cumulant})
is not suitable to describe situations in which the electric field
induces a phase transition; hence, we restrict its application to
relatively small electric fields.
\bibitem{footnote-1}The possible electronic contribution to the EC
effect is not included here since this $H_{\mathrm{eff}}$ does not
incorporate electrons as degrees of freedom. This is probably a small
effect since the major part of the polarization comes from ionic displacements.
\bibitem{dL/dP}For each temperature, the calculations provide $L_{s}$
and $P_{s}$, which allows to plot $L_{s}$ versus $P_{s}$ and then
to obtain the derivative of $L_{s}$ with respect to $P_{s}$ for
any considered temperature. $\frac{dL_{s}}{dP_{s}}$ as a function
of temperature is thus obtained.
\bibitem{note-3} Note that the linear temperature dependence of $a'(T)$
is needed for describing results far away from the Curie temperature,
probably also because the oxygen octahedral tiltings are implicitly
included in our model within some parameters of Eq.\ (\ref{eq:Landau-like-model}).
\bibitem{footnote-2} In case of a ferromagnetic multiferroic, Eq.\ (\ref{eq:Landau-like-model})
needs to be altered by simply replacing the antiferromagnetic vector
by the magnetization, $M$. All the conclusions indicated above thus
still hold but when considering the magnetic Curie temperature rather
than the Néel one and when involving $\frac{\partial M}{\partial P_{s}}$
rather than $\frac{\partial L_{s}}{\partial P_{s}}$.
\end{thebibliography}
\end{document}